\documentstyle[aps]{revtex}
\begin{document}
\bibliographystyle{prsty}
\baselineskip=5.5mm
\parindent=7mm
\begin{center}
{\large {\bf \sc {Calculation of some properties of the vacuum }}} \\[2mm]
Zhi-Gang Wang  \footnote{E-mail,wangzg@mail.ihep.ac.cn}\\[0pt]
{\small  Institute of High Energy Physics, P.O.Box 918 ,Beijing 100039,
P. R. China }\\[0pt]

\end{center}

\begin{abstract}
 In this article, we calculate the dressed quark propagator with the
 flat bottom potential in the framework of the
 rain-bow Schwinger-Dyson equation, which is
 determined by mean field approximation of the global colour model
 lagrangian.  The dressed quark propagator exhibits a dynamical
 symmetry  breaking phenomenon  and gives a constituent quark mass
 about 392 MeV, which is  close to the  value of commonly used
 constituent quark mass in the chiral quark model. Then based on the
   dressed quark propagator, we calculate some properties of the vacuum,
 such as quark condensate, mixed quark condensate
  $g_{s}\langle 0|\bar{q}G_{\mu\nu}\sigma^{\mu\nu}q|0\rangle$,
 four quark condensate
  $\langle 0|\bar{q} \Gamma q\bar{q} \Gamma q |0\rangle$,
 tensor, $\pi$ vacuum susceptibilities.
 The numerical results are compatible with the values of other
 theoretical  approaches.
\end{abstract}


\section{Introduction}
 To investigate the long distance properties of quantum chromodynamics
 (QCD), many approaches are proposed, such as lattice gauge theory,
 QCD sum rules, chiral perturbation theory and phenomenological
 quark models. Each of these approaches has both advantages and
 disadvantages. For example, lattice calculations are rigorous
 from the point of view of QCD, but they suffer from lattice artifacts
 and uncertainties connected with the necessary extrapolation to
 the physical quark masses. The QCD sum rules approach, introduced
 by Shifman, Vainshtein and Zakharov \cite{Shifman,Reinders}
 in 1979, tries to incorporate nonperturbative elements of full
 QCD. As the starting point, operator product expansion (OPE) method
 was used to expand the time ordered currents into a series of quark and
 gluon condensates which parameterize the long distance properties, while
 the short distance effects are incorporated in the Wilson coefficients.
 However, these elements (condensates) can not yet be calculated
 rigorously  in QCD,
 and they are determined from experimental data in the one or other way
 (For example, the quark condensate can be determined
 from experimental data with the help of Gell-Mann-Oakes-Renner
relation \cite{Shifman,Gell}.);
 moreover, although for medium and asymptotic momentum transfers
 the OPE method can be applied for form factors and the moments of wave
 functions \cite{Ioffe1,Nesterenko1,Chernyak1}, at low momentum transfer,
 the standard OPE method cannot be  consistently  applied \footnote{The OPE
 (Wilson expansion) is valid as long as there is a clear distinction between short
 distances ($\propto{ \frac{1}{Q^2}}$) which determine the Wilson coefficients
 and  long distances ($\propto {\frac{1}{\mu^2}}$, $ \mu \sim \Lambda_{QCD}$) which govern the
 condensates. When the  transformed momentum $Q^2$ is small, the
 standard expansion in $ \frac{1}{Q^2}$ would lead to a divergent
 result.}, as pointed out in the early work on photon
 couplings at low momentum for the nucleon  magnetic moments
 \cite{Ioffe,Balitsky}. In Ref.\cite{Ioffe}, the problem was solved by using a
 two point correlator in an external electromagnetic field, with
  vacuum susceptibilities introduced as parameters for
 nonperturbative propagation in the external field. In fact, the existence
 and magnitude of the vacuum susceptibilities are by themselves an important
 physical information about the structure of the vacuum and have many
 applications \cite{Belyaev1,Chiu,Henley2,Henley1}. As
 nonperturbative vacuum properties, the susceptibilities can be
 introduced for both small and large momentum transfers  in the
 external fields .
 In Ref.\cite{Balitsky}, with the long distance effects treated by bilocal
 power corrections and assumption of $\rho$ meson dominance, the authors
 circumvent the problem using a three point formulation. For detailed discussion
  of the relationship between  three point and two point external
  field treatments and the origin of the susceptibilities, one can see
  Ref. \cite{Belyaev2}. The two point  method, however, has two main
  problems: it cannot be used to
 extend  the coupling to medium and high momentum transfer and
 there are additional parameters to be determined, the vacuum
 susceptibilities. In Ref.\cite{Johnson}, by comparing  terms
 appearing in the two point external field expression with those
 in hybrid expansion of the three point function, the authors
 obtain a relationship between the nonperturbative elements
 in the two methods. From these relationships, one can
 express  the induced susceptibilities of the
 two point method in terms of well-defined four quark vacuum
 matrix elements. These susceptibilities may play an important role
 in the QCD sum rules approach. In particular, the strong and
 parity-violating  pion-nucleon  coupling depends crucially on
 the $\pi$ vacuum  susceptibility ($\chi^{\pi}a$),
   while the tensor vacuum susceptibilities
   are relevant for the determination of the tensor
  charge of the nucleon, which is connected through deep-inelastic
   sum rules to the leading-twist nucleon structure functions and
    transversity distribution \cite{He1,He2,Ralston}.
  Although in principle the pion
 susceptibility can be estimated using PCAC, in practice there
 are inconsistencies with the values needed phenomenologically
 \cite{Johnson,Henley1}. On the other hand, as it has been pointed out in
  Ref.\cite{Belyaev}, that treatment of the vacuum tensor
  susceptibility is subtle and different treatments can lead to
  very different results for the tensor charge.
  Therefore, it is interesting to calculate these vacuum condensates and
  vacuum susceptibilities within
 the framework of coupled flat-bottom potential (FBP)
 model \cite{Wang} and
 global colour model (GCM) \cite{Cahill,Frank,Roberts,Tandy}.

The global colour model (GCM), a NJL-like theory \cite{Nambu},
has provided a lot of successful descriptions of the long distance properties of strong interaction  and the QCD
 vacuum as well as hadronic phenomena at low energy based on the
 theoretical foundation that the quark propagator contains valuable information about
the nonperturbative properties of QCD \cite{Cahill,Frank,Roberts,Tandy}. The  flat-bottom potential is a sum of Yukawa potentials, which
  not only  satisfies gauge invariance, chiral invariance and fully
relativistic covariance, but also suppresses the singular point which the
 Yukawa potential has. It works well in understanding the meson
 structure, such as electromagnetic form factor, radius, decay constant
  \cite{Wang}.

 In this article, we combine  the GCM with the FBP
 to calculate the quark condensate,
 $g_{s}\langle 0|\bar{q}G_{\mu\nu}\sigma^{\mu\nu}q |0\rangle$,
 $\langle 0|\bar{q} \Gamma q\bar{q} \Gamma q |0\rangle$ ,
 tensor, pion vacuum susceptibilities in the framework of the rain-bow
 Schwinger-Dyson (SD) equation. The article is arranged as follows:
 in section 2, we brief out line the GCM and in section 3,
  introduce the FBP, SD equation; in  section 4, we calculate
  the vacuum condensates and vacuum
 susceptibilities; in section 5, conclusion and discussion.

 \section{Global Colour Model}

Here we brief out line the main skeleton of the GCM.
The global colour model, based upon an effective quark-quark interaction which
is approximated by the effective flat bottom potential in this article,
is defined through a truncation of QCD lagrangian while maintaining
all global symmetries.

The generating function for QCD in Euclidean space can be written as
\begin{equation}
Z[ \bar{\eta}, \eta] = \int {\cal{D}}\bar{q} {\cal{D}} q
{\cal{D}} A \exp\left\{-S
+ \int d^4 x (\bar{\eta} q + \bar{q} \eta)\right\} \,
\end{equation}
where
$$ S = \int d ^4 x \left\{ \bar{q} \left[ \gamma _{\mu}
\left(\partial_{\mu}
- i g \frac{\lambda ^{a}}{2} A^{a}_{\mu} \right)+m \right] q +
\frac{1}{4}
G_{\mu \nu}^{a} G_{\mu \nu}^{a} \right\}, $$
and $G_{\mu \nu}^{a}=\partial_{\mu}A^{a}_{\nu}-\partial_{\nu}
A^{a}_{\mu}+gf^{abc}A^{b}_{\mu}A^{c}_{\nu}$.
We leave the gauge fixing term, the ghost field term and its
integration measure to be understood.
Here we introduce
\begin{equation}
  e^{W[J]}=\int{\cal D}Ae^{\int d^{4}x(-{1\over
  4}G^a_{\mu\nu}G^a_{\mu\nu}+J^a_{\mu}A^a_{\mu})}.
\end{equation}
So the generating function can be written as
\begin{equation}
  Z[ \bar{\eta}, \eta]=\int{\cal D}q{\cal D}\bar{q}e^{-\int d^{4}x
  [\bar{q}({\gamma \cdot \partial}+m)q -\bar{\eta}q-\bar{q}\eta]}
          e^{W[J]}.
\end{equation}

The functional $W[J]$ can be formally expanded in the current
$J^a_{\mu}$:
\begin{equation}
W[J]={1\over 2}\int d^{4}xd^{4}y J^a_{\mu}(x)D^{ab}_{\mu\nu}
(x,y)J^b_{\nu}(y)
+\frac{1}{3!}\int J^a_{\mu}J^b_{\nu}J^c_{\rho}D^{abc}_{\mu\nu\rho}+\cdots.
\end{equation}
The GCM is defined through a truncation of the functional $W[J]$ in which
the higher order $n(\ge3)$-point functions are neglected, and only the
gluon
2-point function $D^{ab}_{\mu\nu}(x,y)$ is retained. This model maintains all
global
symmetries of QCD and permits a $1/N_c$ expansion, however
local SU(3) gauge invariance is lost by the truncation.

Using the functional integration approach, the generating function of
the truncation is given by
\begin{eqnarray}
  Z[ \bar{\eta}, \eta]_{\rm GCM}=\int{\cal D}q{\cal D}\bar{q}\exp
  \large{(} -\int d^{4}x \bar{q}( \gamma \cdot \partial+m)q
   -\frac{g^2}{2}\int d^{4}xd^{4}yj^a_{\mu}(x)
 D^{ab}_{\mu\nu}(x-y)j^b_{\nu}(y)
  +\int d^{4}x (\bar{\eta} q + \bar{q}\eta ) \large{)},
\end{eqnarray}
where $j^a_{\mu}(x)=\bar{q}(x)\gamma_{\mu}\frac{\lambda^a}{2}q(x)$ is the
quark   color current.  For
simplicity we use a Feynman-like gauge
$D^{ab}_{\mu\nu}(x-y)=\delta_{\mu\nu}\delta^{ab}D(x-y)$. Performing  the standard
bosonization procedure, (For detailed discussion of this topic,
see Ref. \cite{Tandy}) we obtain the resulting expression for the
generating function in
terms of the bilocal field integration
\begin{equation}
Z_{\rm GCM}=\int{\cal
D}{\cal B}e^{-S[{\cal B}]} ,
\end{equation}
 where the action is given by
\begin{equation}
 S[{\cal B}]=-{\rm TrLn}[G^{-1}]+\int d^{4}xd^{4}y
  \frac{{\cal B}^{\theta}(x,y){\cal B}^{\theta}(y,x)}{2g^2D(x-y)}
  +\int d^{4}xd^{4}y \bar{\eta}(x) G(x,y;[{\cal B}^{\theta}])\eta(y)
  ,
\end{equation}
and the inverse quark's  Green function $G^{-1}$ is defined as
\begin{equation}
 G^{-1}(x,y)= \gamma\cdot\partial \delta(x-y)+\Lambda^{\theta}{\cal B}^{\theta}(x,y).
\end{equation}
Here the quantity $\Lambda^{\theta}$ arises from Fierz reordering of the
current-current interaction term in Eq.(5),
\begin{equation}
 \Lambda^{\theta}_{in}\Lambda^{\theta}_{mj}=
 (\gamma_{\mu}\frac{\lambda^a}{2})_{ij}
 (\gamma_{\mu}\frac{\lambda^a}{2})_{mn}
\end{equation}
and is the direct product of Dirac, flavor SU(3) and color matrices,
\begin{equation}
  \Lambda^{\theta}={1\over 2}(1_D,i\gamma_5,\frac{i}{\sqrt{2}}\gamma_{\mu},
  \frac{i}{\sqrt{2}}\gamma_{\mu}\gamma_5)\otimes(\frac{1}{\sqrt{3}}1_F,
  \frac{1}{\sqrt{2}}\lambda^a_F)\otimes({4\over 3}1_c,
  \frac{i}{\sqrt{3}}\lambda^a_c).
\end{equation}

The vacuum configurations are defined by minimizing the bilocal action:
$\left. \frac{\delta S[{\cal B}]}{\delta {\cal B}}
 \right |_{\bar{\eta} , \eta=0}=0,$
following this mean field approximation, we obtain
\begin{equation}
  {\cal B}^{\theta}_0(x-y)=g^2D(x-y){\rm tr}[\Lambda^{\theta}G_0(x-y)].
\end{equation}
These configurations provide self-energy dressing of the quarks through the
definition
$\Sigma(p)\equiv\Lambda^{\theta}{\cal B}^{\theta}_0(p)=i \gamma \cdot p[A(p^2)-1]
+B(p^2)$. In fact, equation (11) is the rain-bow SD equation.
In terms of $A$ and $B$, the quark's Green function at
${\cal B}^{\theta}_0$
is given by
\begin{equation}
  G_0(x,y)=G_0(x-y)=\int\frac{{\rm d}^4p}{(2\pi)^4}
  \frac{-i \gamma \cdot p A(p^2)+B(p^2)}{p^2A^2(p^2)+B^2(p^2)}e^{ip\cdot(x-y)}.
\end{equation}

With the  GCM generating function Eq.(6), it is straightforward
to calculate the vacuum expectation value of any quark operator of the
form
\begin{equation}
O_{n}\equiv(\bar{q}_{i_{1}}\Lambda^{(1)}_{i_{1}j_{1}}q_{j_{1}})
(\bar{q}_{i_{2}}\Lambda^{(2)}_{i_{2}j_{2}}q_{j_{2}})\cdots
(\bar{q}_{i_{n}}\Lambda^{(n)}_{i_{n}j_{n}}q_{j_{n}}),
\end{equation}
in the mean field vacuum.
Here the $\Lambda^{(i)}$ represents an operator  in
Dirac, flavor and color space.

Taking the appropriate number of derivatives with respect to
the external sources
$\eta_{i}$ and $\bar{\eta_{j}}$ of Eq.(6)
and putting $\eta_{i}=\bar{\eta}_{j}=0$ \cite{Negele},
we obtain
\begin{equation}
\langle 0|:O_{n}:|0\rangle=(-)^{n}\sum_{p}[(-)^{p}\Lambda^{(1)}_{i_{1}j_{1}}
\cdots\Lambda^{(n)}_{i_{n}j_{n}}(G_0)_{j_{1}i_{p(1)}}\cdots(G_0)_{j_{n}i_{p(n)}}],
\end{equation}
where p stands for a permutation of the n indices.
 From Eq.(14), we can easily obtain the expression for the nonlocal
 four quark condensate, which is another important vacuum condensate in
 the QCD sum rules approach  beside quark condensate.
   \begin{eqnarray}
   \langle 0|:\bar{q}(x)\Lambda^{(1)}q(x)\bar{q}(y)
   \Lambda^{(2)}q(y):|0\rangle
   =-tr_{\gamma C}[G_{0}(y,x)\Lambda^{(1)}
   G_{0}(x,y)\Lambda^{(2)}]
   +tr_{\gamma C}[G_0(x,x)\Lambda^{(1)}]
   tr_{\gamma C}[G_{0}(y,y)\Lambda^{(2)}].
   \end{eqnarray}

\section{Flat Bottom Potential and Schwinger-Dyson Equation}
The phenomenological FBP is a sum of Yukawa potentials which is an analogous to the
exchange of a series of particles and ghosts with different
masses,
\begin{equation}
G(k^{2})=\sum_{j=0}^{n}
 \frac{a_{j}}{k^{2}+(N+j \rho)^{2}}  ,
\end{equation}
where $N$ stands for the minimum value of the mass, $\rho$ is their mass
difference, and $a_{j}$ is their relative coupling constant.\
In its three dimensional form, the FBP takes the following form:
\begin{equation}
V(r)=-\sum_{j=0}^{n}a_{j}\frac{{\rm e}^{-(N+j \rho)r}}{r}  .
\end{equation}
In order to suppress the singular point at $r=0$, we take the
following conditions:
\begin{eqnarray}
V(0)=constant, \nonumber \\
\frac{dV(0)}{dr}=\frac{d^{2}V(0)}{dr^{2}}=\cdot \cdot \cdot=\frac{d^{n}V(0)}
{dr^{n}}=0    .
\end{eqnarray}
So we can determine $a_{j}$ by solve the following
equations, inferred from the flat bottom condition Eq.(18),
\begin{eqnarray}
\sum_{j=0}^{n}a_{j}=0,\ \
\sum_{j=0}^{n}a_{j}(N+j \rho)=V(0), \ \
\sum_{j=0}^{n}a_{j}(N+j \rho)^{2}=0,\ \
\cdots \ \
\sum_{j=0}^{n}a_{j}(N+j \rho)^{n}=0 .
\end{eqnarray}
As in  previous literature \cite{Wang}, $n$ is set to be
9. In Ref. \cite{Wang} (Phys. Rev. {\bf D47} (1993) 2098),
the authors solve the Bethe-Salpeter equation based on expansion
in terms of Gegenbauer polynomials with FBP, then use the
Bethe-Salpeter amplitudes as input to calculate the $\pi$ meson
electromagnetic form factor. Among many other values, the
phenomenological value 9 can give stable solution and best fit to
experimental data\footnote{Private communication.}.

In the rainbow approximation, the SD equation takes the following form:
\begin{equation}
S^{-1}(p)=\gamma \cdot p - \hat{m}+\frac{16 \pi i}{3}\int \frac {d^{4}k}{(2 \pi)^{4}} \gamma_{\mu}
S(k)\gamma_{\nu}G^{\mu \nu}(k-p),
\end{equation}
where
\begin{eqnarray}
S^{-1}(p)&=& A(p^2)\gamma \cdot p-B(p^2)\equiv A(p^2)
[\gamma \cdot p-m(p^2)], \\
G^{\mu \nu }(k)&=&-g^{\mu \nu}G(k^2),
\end{eqnarray}
and $\hat{m}$ stands for an explicit quark mass-breaking term.
 With the explicit small mass term, we can preclude the zero
 solution for $B(p)$ and in fact there indeed exist a bare current
 quark mass. In order to keep consistently with section two,
here we take Feynman gauge.
This dressing comprises the notation of constituent quarks by
 providing a mass $m(p^2)=B(p^2)/A(p^2)$, which is corresponding to
 dynamical symmetry breaking. Because the form of
 the gluon propagator $g^2G(p)$ in the infrared region is unknown,
 one often uses model forms as input in the previous studies
  of the rainbow SD equation \cite{Cahill,Roberts,Tandy,Meissner}.

  As a nonperturbative phenomenological potential, the FBP may have
intrinsic connections with other nonperturbative phenomena, such as
confinement, instanton, vacuum condensate and renormalon, glueball.
 In this article, we concern  mainly the quark condensates, while the
vacuum condensates play an important role in determining the
properties of the mesons, glueballs and their mixing as indicated
by the low energy theorem \cite{Novikov}. However, the realization
of gluonic freedom in hadronic models is more ambiguous and
difficulty then that for quarks, the glueball spectra obtained from
  two main nonperturbative methods
 (lattice QCD and QCD sum rules )
  have some contradictions \cite{West}. Here we just replace the
  gluon propagator with FBP and suppose the nonperturbative
  effects are incorporated in it, the possible effects on glueball
  will be our next work.
 \footnote{ For example, we can obtain the gluon condensate from the
nonperturbative gluon propagator (the effective potential),
then based on the gluon condensate and low energy theorem, we can
study the lowest lying glueball mass, their mixing, and so on.
There is another way to study the flat bottom potential to glueball
spectrum through the Bethe-Salpeter equation.}
  Although there may be debates against
  doing that, the quark propagator obtained with the FBP does lead to
  satisfactory dynamical symmetry breaking phenomenon and quark confinement
  based on the theoretical foundation that there no singularities
  on the real timelike $p^2$ axial \cite{Wangwan}.

 In this article,
we assume that a Wick rotation to Euclidean variables is
allowed, and perform a rotation analytically continuing $p$ and $k$ into the
Euclidean region where them can be denoted by $\bar{p}$ and $\bar{k}$, respectively.
The Euclidean SD equation can be projected into two coupled integral
equations for $A(\bar{p}^2)$ and $B(\bar{p}^2)$. For simplicity, here
ignore the bar on $p$ and $k$.
 Numerical values for $A(p^2)$, $B(p^2)$ and $m(p^2)$
are shown in Fig.[1] .\\

\section{Vacuum Condensates and Vacuum Susceptibilities}
In this section, we calculate
 the quark condensate
  $\langle 0|\bar{q}q |0\rangle$, the mixed quark gluon condensate
$g_{s}\langle 0|\bar{q}G_{\mu\nu}\sigma^{\mu\nu}q |0\rangle$, the four
quark condensate $\langle 0|\bar{q} \Gamma q\bar{q} \Gamma q |0\rangle$
and tensor, pion vacuum susceptibilities. The numerical results
are compared with other theoretical works.

Here we take a short digression to discussing gauge invariance.
The propagation of a colored quark in the background gluon field
results in a phase factor given by the exponential of a line
integral,
\begin{equation}
\hat{P}(x,y;A,C)=P exp \{ ig \int_{C} A_{\mu}(z) dz^{\mu}\},
\end{equation}
where $P$ is a path order operator, stands for that the color
matrices should be ordered along the path $C$ connecting the point
$x$ and $y$. The gauge dependence of the quark propagator can be
factored out as an ''A-dependent'' phase
before taking the vacuum expectation value. In operator expanding
 of the correlators of colorless currents, these phase factors
 can be canceled out with each other,  one can facilitate the
 calculation by imposing Fock-Schwinger gauge
 condition \cite{Nikolaev1,Nikolaev2},
 \begin{equation}
 (x^{\mu}-x_{0}^{\mu})A_{\mu}(x)=0,
 \end{equation}
or in the other word, let $\hat{P}=1$.  To gauge
dependent quantities, (for example, $\langle 0| \bar{q}(x)q(0) |0\rangle
$), we can insert phase factors $\hat{P}$ to produce the
corresponding gauge invariant ones ( $\langle 0| \bar{q}(x) \hat{P}(x,0;A,C)q(0)
|0\rangle$). The gauge invariant quantity
$\langle 0|: \bar{q}(x) \hat{P}(x,0;A,C) q(0): |0 \rangle$
calculated in a special gauge, for example, Fock-Schwinger gauge, will not
change its value. In the following, the gauge dependent quantities
can be taken as their corresponding invariant ones in the special gauge,
Fock-Schwinger gauge.
 Furthermore, although the gauge
dependent quantities, such as  $\langle 0|: \bar{q}(x) q(0): |0
\rangle$,
 are dependent on the special gauge we choose, they are important
 imformations in describing the nonperturbative QCD vacuum. In a sense,
 the gauge dependence is fictitious, since the vacuum field, after
 averaging, is clearly invariant under transformations.

\subsection{Vacuum Condensates}
 The quark propagator is defined as
\begin{equation}
S(x)=\langle 0 |T[q(x) \bar q(0)]|0 \rangle  .
\end{equation}
where q(x) is the quark field and T the time-ordering operator.
For the physical vacuum consists of both perturbative and nonperturbative
parts, so the quark propagator S(x) can be divided into a perturbative
and a nonperturbative part as the following:
\begin{eqnarray}
S(x)= S_{PT}(x)+S_{NP}(x) .
\end{eqnarray}
In the nonperturbative vacuum, the normal-ordered product $S_{NP}(x)$
does not vanish. For short distance, the OPE for the scalar part of $S_{NP}(x)$
gives
\begin{equation}
\langle0|: \bar q(x) q(0) :|0 \rangle=\langle 0|: \bar q(0) q(0): |0 \rangle
- \frac{x^2}{4}(\langle 0| : \bar q(0) \sigma \cdot G(0)  q(0) :|0 \rangle
+ \cdots
\end{equation}
in which the local operators of the expansion are the quark condensate,
the mixed condensate, and so forth.
In Ref.\cite{Kisslinger2} the nonlocal condensate is put in the
following form:
\begin{equation}
\langle 0|: \bar q(x) q(0) :|0 \rangle=g(x^2)\langle0| : \bar q(0) q(0) :|0 \rangle
,
\end{equation}
where $g(x^2)$ is the vacuum non-locality of the nonlocal quark
condensate.
The nonlocal quark condensate  $\langle0| : \bar q(x) q(0) :|0 \rangle$ is given
then by the scalar part of the Fourier transformed inverse quark
propagator, which can be inferred from Eq.(14).
\begin{eqnarray}
\langle0|:\bar{q}(x)q(0):|0\rangle&=&(-)tr_{\gamma C}[G_{0}(x,0)]
\nonumber\\
&=&(-4N_{c})\int^{\mu}_{0}\frac{d^4 p}{(2\pi)^4}\frac{B(p^2)}
{p^2A^2(p^2)+B^2(p^2)}e^{ipx}\nonumber\\
&=&(-)\frac{12}{16\pi^2}\int^{\mu}_{0}ds s\frac{B(s)}
{sA^2(s)+B^2(s)}[2\frac{J_1(\sqrt{sx^2})}{\sqrt{sx^2}}].
\end{eqnarray}

At x=0 the expression for the local condensate $ \langle 0|: \bar q(0) q(0) :|0 \rangle $
is recovered,
\begin{equation}
\langle 0|: \bar q(0) q(0) :|0 \rangle = -\frac{12}{16 \pi^2}
\int_{0}^{ \mu} d s s\frac {B(s)}{sA^2(s)+B^2(s)}.
\end{equation}
The non-locality $g(x^2)$ can be obtained immediately by dividing
Eq.(29) through Eq.(30).
Our analysis ignores effects from hard gluonic radiative correction to
the condensates which are connected to a possible change of the renormalization scale
$\mu$ at which the condensates are defined. As in Ref.\cite{Meissner},
$\mu$ is taken to be 1 $GeV^2$.
 Final we obtain the nonlocal quark condensate which shown in
 Fig.[2] and local quark condensate which shown in Table 1.
 From Fig.[2], we can see that the nonlocality is  compatible with other theoretical
 calculations, including instanton liquid model \cite{Dorokhov},
 effective quark-quark
  interaction model \cite{Kisslinger2}, GCM \cite{Zong} ,
  dipole fit approach \cite{Johnson}, and other potential model \cite{Wang5}.
  The curves obtained in Ref.\cite{Kisslinger2} and in Ref. \cite{Johnson}
  are similar to each other, here we choose the curve in Ref.\cite{Kisslinger2}
  to compare with our result. For simplicity, in
  Fig.2, we list three curves.

  In \cite{Wanwan}, we calculate the nonlocality of the quark
   condensate; in calculation, we find that there are other parameters
   can give correct nonlocality curve. In the present work, we re-fit
   the parameters to incorporate the mixed condensate and the vacuum
   susceptibilities. Furthermore, in the present work, we change the
   coefficient of FBP inserted in the SD equation by a factor $\pi$,
   the present formulation is better than the old one \cite{Wanwan,Wang}.

   By means of Eq.(15), one can calculate all kinds of nonlocal four quark
    condensates . If we take $\Lambda^{(1)}$
   =$\Lambda^{(2)}$=$\gamma_{\mu}\frac{\lambda^a_{C}}{2}$ then
   \begin{eqnarray}
   & &\langle 0|:\bar{q}(x)\gamma_{\mu}\frac{\lambda^a_{C}}{2}q(x)\bar{q}(0)
   \gamma_{\mu}\frac{\lambda^a_{C}}{2}q(0):|0\rangle\nonumber\\
   &=&(-)\int^{\mu}_{0}\int^{\mu}_{0}
   \frac{d^4p}{(2\pi)^4}\frac{d^4q}{(2\pi)^4}e^{ix\cdot(p-q)}
   \left[4^3\frac{B(p^2)}{A^2(p^2)p^2+B^2(p^2)}\frac{B(q^2)}
   {A^2(q^2)q^2+B^2(q^2)}\right.\nonumber\\
   & &\left.+ 2\times 4^2\frac{A(p^2)}{A^2(p^2)p^2+B^2(p^2)}\frac{A(q^2)}
   {A^2(q^2)q^2+B^2(q^2)}p\cdot q\right].
   \end{eqnarray}
   Similarly, at x=0 we obtain the expression for the local four quark condensate
   $\langle0|:\bar{q}\gamma_{\mu}\frac{\lambda^a_{C}}{2}q\bar{q}
   \gamma_{\mu}\frac{\lambda^a_{C}}{2}q:|0\rangle$,
   \begin{eqnarray}
   \langle0|:\bar{q}\gamma_{\mu}\frac{\lambda^a_{C}}{2}q\bar{q}
   \gamma_{\mu}\frac{\lambda^a_{C}}{2}q:|0\rangle =(-4^3)[\int^{\mu}_{0}\frac{d^4p}{(2\pi)^4}
   \frac{B(p^2)}{A^2(p^2)p^2+B^2(p^2)}]^2
    =(-)\frac{4}{9}\langle0|:\bar{q}q:|0\rangle^2 .
   \end{eqnarray}

   As far as the mixed condensate
   $g_{s}\langle\bar{q}G_{\mu\nu}\sigma^{\mu\nu}q\rangle$ is concerned , the
   evaluation is somewhat lengthy, one
   can use the method described by Ref.\cite{Meissner} to obtain the mixed condensate in
   Euclidean space,
   \begin{eqnarray}
   g_{s}\langle 0|\bar{q}G_{\mu\nu}\sigma^{\mu\nu}q |0\rangle
   =(-)(\frac{N_{c}}{16\pi^2}) ( \frac{27}{4}
   \int_{0}^{\mu}dss\frac{B[2A(A-1)s+B^2]}
   {A^2s+B^2} +12\int_{0}^{\mu}dss^2\frac{B(2-A)}{A^2s+B^2} ).
   \end{eqnarray}

     In Table 1. we display our results for $\langle\bar{q}q\rangle$ and
   $g_{s}\langle\bar{q}G_{\mu\nu}\sigma^{\mu\nu}q\rangle$  and compare
   them with the corresponding values  obtained from  other
   theoretical approaches, such as
   QCD sum rules \cite{Narison}, quenched lattice QCD \cite{Kremer},
   the instanton liquid model \cite{Polyakov}
   , the model of confining gluon propagator for GCM \cite{Meissner} and
    other potential model \cite{Zong1,Yang,Wang5} .
   From Table 1, we can see that the values  are compatible with other
   theoretical works.
\subsection{Vacuum Susceptibilities}

In the external field of QCD sum rule two--point method, one often
 encounters the quark propagator in the presence of a external
 current
$J^{\Gamma}(y)=\bar{q}(y)\Gamma q(y)$($\Gamma$ stands for
 the appropriate combination of Dirac , flavor and colour matrices).
\begin{eqnarray}
S^{cc'\Gamma}_{\alpha\beta}(x)=\langle 0|T[q^{c}_{\alpha}(x)
\bar{q}^{c'}_{\beta}(o)]0
\rangle_{J^{\Gamma}}
=S^{cc'\Gamma,PT}_{\alpha\beta}(x)
+S^{cc'\Gamma,NP}_{\alpha\beta}(x),
\end{eqnarray}
where $S^{c c'\Gamma,PT}_{q}(x)$ is the quark propagator coupled
perturbatively
to the current and $S^{c c'\Gamma,NP}_{q}(x)$ is the nonperturbative quark
propagator in the presence of the external current $J^{\Gamma}$
(To be explicitly, here we keep all the indexes).

The vacuum susceptibility $\chi^{\Gamma}$ in the QCD sum rule two--point
external field treatment can be defined as \cite{Johnson}
\begin{eqnarray}
S^{cc'\Gamma,NP}_{\alpha\beta}(x)&=&\langle 0|:q^{c}_{\alpha}(x)
\bar{q}^{c'}_{\beta}(0):|0\rangle_{J^{\Gamma}}
=-\frac{1}{12}\Gamma_{\alpha\beta}\delta_{cc'}
\chi^{\Gamma}H(x)\langle 0|:\bar{q}(0)q(0):|0\rangle ,
\end{eqnarray}
where the phenomenological function $H(x)$ represents the nonlocality of
 the
two quark nonlocal condensate. Obviously H(0)=1.

By  comparing  terms
 appearing in the two point external field expression with those
 in hybrid expansion of the three point function, one can
 obtain a relationship between the nonperturbative elements
 in the two methods. From these relationship, one can
   express the induced susceptibilities of the
 two point method in terms of well-defined four quark vacuum
 matrix element \cite{Johnson}.

The presence of external field implies that
$S^{cc'\Gamma}_{\alpha\beta}(x)$
is evaluated with an additional term
$\Delta L\equiv -J^{\Gamma}\cdot\phi_{\Gamma}$
added to the usual QCD Lagrangian, where $\phi_{\Gamma}$ is the value of
external field. In three point method of QCD sum rule,
if one takes only a linear external
field approximation, the $S^{cc'\Gamma,NP}_{\alpha\beta}(x)$ in
Euclidean space
is given by
\begin{equation}
S^{cc'\Gamma,NP}_{\alpha\beta}(x)=
\int d^{4}y~e^{-iq\cdot y}\langle 0|:q^{c}_{\alpha}(x)\bar{q}(y)
\Gamma q(y)\bar{q}^{c'}_{\beta}(0):|0\rangle .
\end{equation}

Using Eq.(35) and Eq.(36) we obtain
\begin{equation}
-\frac{1}{12}\delta_{cc'}\Gamma_{\alpha\beta}\chi^{\Gamma}
\langle 0|:\bar{q}(0) q(0):|0 \rangle
=\int d^{4}y~e^{-iq\cdot y}\langle 0|:q^{c}_{\alpha}(0)\bar{q}(y)
\Gamma q(y)\bar{q}^{c'}_{\beta}(0):|0\rangle .
\end{equation}

Multiplying Eq.(37) by $\Gamma_{\beta\alpha}\delta_{cc'}$, we get
the result:
\begin{equation}
\chi^{\Gamma}a
=-\frac{16\pi^2}{tr_{\gamma}(\Gamma\Gamma)}\int d^{4}y~ e^{-iq\cdot y}
\langle 0|:\bar{q}(0)\Gamma q(0)
\bar{q}(y)\Gamma q(y):|0\rangle .
\end{equation}
In the case of tensor current($\Gamma=\sigma_{\mu\nu}$),
according to Eq.(14), we calculate the tensor matrix element
explicitly,
\begin{eqnarray}
 & &\langle0|\bar{q}(0)\sigma_{\mu\nu}q(0)\bar{q}(y)\sigma_{\mu\nu}
 q(y)
             |0\rangle  \nonumber \\
 &=&{\rm tr}_{\gamma c}\int\frac{{\rm d}^4p}{(2\pi)^4}\sigma_{\mu\nu}
     \frac{-i \gamma \cdot p A(p^2)+B(p^2)}{A(p^2)p^2+B(p^2)}
     {\rm tr}_{\gamma c}\int\frac{{\rm d}^4k}{(2\pi)^4}\sigma_{\mu\nu}
     \frac{-i\gamma \cdot k A(k^2)+B(k^2)}{A(k^2)k^2+B(k^2)} \nonumber \\
 & &{}-\int\int\frac{{\rm d}^4p}{(2\pi)^4}\frac{{\rm d}^4k}{(2\pi)^4}
     e^{-i(p-k)\cdot y}{\rm tr}_{\gamma c}\left [\sigma_{\mu\nu}
    \frac{-i\gamma \cdot p A(p^2)+B(p^2)}{A(p^2)p^2+B(p^2)}\sigma_{\mu\nu}
    \frac{-i\gamma \cdot k A(k^2)+B(k^2)}{A(k^2)k^2+B(k^2)}\right ]   \nonumber \\
 &=&-48N_c\int\int\frac{{\rm d}^4p}{(2\pi)^4}\frac{{\rm d}^4k}{(2\pi)^4}
     e^{-i(p-k)\cdot y}\frac{B(p^2)}{A(p^2)p^2+B(p^2)}\frac{B(k^2)}{A(k^2)k^2+B(k^2)},
\end{eqnarray}

where $N_c=3$ is the number of colors.

In the limit~$q\rightarrow0$, we obtain

\begin{eqnarray}
\chi^{T}a
=3\int^{\mu}_{0}sds\left[\frac{B(s)}{A^2s+B^2(s)}\right]^2 =0.0757GeV^2.
\end{eqnarray}

From Eq.(36) we can see that the vacuum susceptibility originates from the nonlocal four quark
  condensate contribution.
  In Ref.\cite{Bakulev}, the authors get the same conclusion based on
  completely different  viewpoint of duality.

  The vacuum susceptibility $\chi^{T}a$ defined in Eq.(38) (same as Ref. \cite{Zong1}) has  an opposite sign and a factor $4 \pi^2$ larger
 than the definition in Refs.\cite{He1,Belyaev,Bakulev}. In order to compare
  with those  estimations, we make a redefinition
\begin{equation}
\chi^{T}a \rightarrow  -\frac{\chi^{T}a}{4 \pi^2} =-0.0019 GeV^2 .
\end{equation}

In Table 2, we compare the result with the values of other theoretical
estimations of the vacuum tensor susceptibility.

In the following, we calculate the $\pi$  vacuum susceptibility
which is crucial to determine the strong and
party-violating pion-nucleon coupling. In the case of pseudoscalar
current, we take $\Lambda^{(1)}=\Lambda^{(2)}=\gamma^{5}$ in Eq.(15),
and get the $\pi$ vacuum susceptibility $\chi^{\pi}a$:
\begin{eqnarray}
&&\chi^{\pi}a
=12\pi^2\int d^4y~tr_{\gamma}\left[
\int \frac{d^4p}{(2\pi)^4}\frac{-i\gamma\cdot p A+B(p^2)}{A^2p^2+B^2(p^2)}
e^{-ip\cdot y}\gamma_{5}
\int \frac{d^4q}{(2\pi)^4}\frac{-i\gamma\cdot q A+B(q^2)}{A^2q^2+B^2(q^2)}
e^{iq\cdot y}\gamma_{5}\right] \nonumber \\
&&=3\int^{\mu}_{0}sds\frac{1}{A^2s+B^2(s)}=1.06 ~GeV^2,
\end{eqnarray}
which is below  the range $\chi^{\pi}a\simeq (1.7-3.0)~GeV^2$
obtained within a phenomenological approach \cite{Johnson}. In
Ref.\cite{Johnson}, the author made a simple estimation  of the value
based on  the space time structure of the nonlocal quark
 condensate extracted from experimental data on sea-quark distributions but
 with a crude assumption of  vacuum saturation for the intermediate states.

\section{Conclusion and Discussion}

 In this article, we calculate the dressed quark propagator with the
 FBP in the framework of the rain-bow SD equation, which is
 determined by mean field approximation of the GCM
 lagrangian.  The dressed quark propagator exhibits a dynamical symmetry
 breaking phenomenon
 and gives a constituent quark mass about 392 MeV, which is  close to the
  value of commonly used constituent quark mass 350 MeV in the chiral quark model. Then based on the
   dressed quark propagator, we  obtain the  quark condensate
   $\langle 0|\bar{q}q|0\rangle$, the mixed quark
 gluon condensate
$g_{s}\langle0|\bar{q}G_{\mu\nu}\sigma^{\mu\nu}q|0\rangle$, the four
quark condensate $\langle0|\bar{q} \Gamma q\bar{q} \Gamma q|0\rangle$  and
tensor, pion vacuum susceptibilities at the mean field level, which
are compared with other theoretical  calculations.  The values
of the nonlocality of quark condensate, local quark condensate, mixed
quark condensate are compatible with other theoretical results.
While the value of $\pi$ susceptibility is below that estimated in
Ref.\cite{Johnson}, which is based on the space-time structure of
the nonlocal quark condensate and a crude assumption of vacuum
saturation for the intermediate states.       In Table 3, we also
compare with other theoretical results.
  In Table 2, we can see that  the numerical value of
 the tensor vacuum susceptibility  varies with theoretical
 approach.
  It  has been pointed out
  in Ref.\cite{He1} that experimental measurement of the
  tensor charge of the nucleon
  is possible, and one might be able  to test the theoretical
   predictions of the tensor susceptibility in the future.

In solving the rain-bow SD equation, the parameters are taken as
$\Lambda_{QCD}=200MeV$,$N=3.0 \Lambda_{QCD}$, $V(0)=-12.0\Lambda_{QCD}$,
$\rho=2.0\Lambda_{QCD}$, $m_{u}=m_{d}=8 MeV$ and the large momentum
cut-off $L=630(\Lambda_{QCD})$.

In calculation, the coupled integral equations for the quark
propagator functions $A(p^{2})$ and $B(p^{2})$
are solved numerically by simultaneous iterations. The iterations converge
rapidly to a unique stable solution of propagator functions and independent
the initial guesses. The propagator functions $A(p^{2})$ and
$B(p^{2})$ are shown in Fig.[1] , at small $p^{2}$ , $A(p^{2})$ differs from
the value 1 appreciably, while it tends to 1 for large $p^{2}$.
We find that at small $p^{2}$ , $m(p^{2})$ is greatly re-normalized,
 while at large $p^{2}$,
it takes asymptotic behaviour. For $u$ and $d$ quark, $m(0)=392 MeV$,
which is close to the constituent quark masses, the
connection of $m(p)$ to constituent masses is somewhat less direct
and is precise only for heavy quarks. For heavy quarks,
$ m_{constituent}(p)=m(p=2m_{constituent}(p))$ , for light quarks ,
it only makes a crude estimation \cite{Politzer}. At about $p=1GeV$,
the mass function grows rapidly as the momentum decreases, that is
an indication of dynamical symmetry breaking.

The results show that the phenomenological FBP works well with the
rainbow SD equation in calculating nonlocal quark vacuum condensate, mixed condensate
and vacuum susceptibilities.
This can be extend to other system, such as $q$$\bar{q}$ system, $\bar{q}$$Q$ system, vector meson and $qqq$ system or by applying the results to the
calculation of quantities and processes requiring detailed
knowledge of the quark propagator .
\begin{center}
\bf{ Acknowledgement}
\end {center}

The author would like to
thank Prof. T.Huang
 for his helpful discussion.

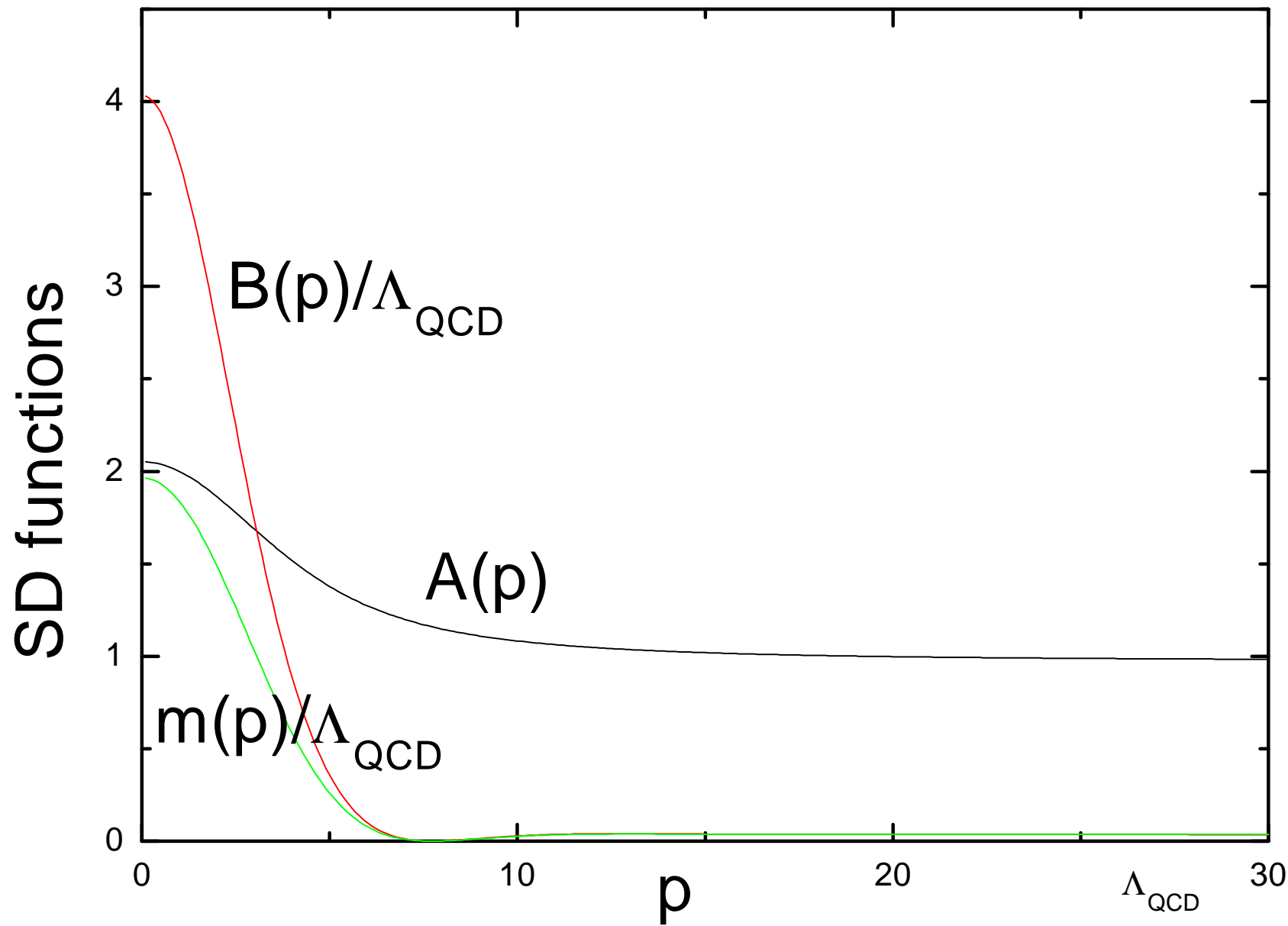
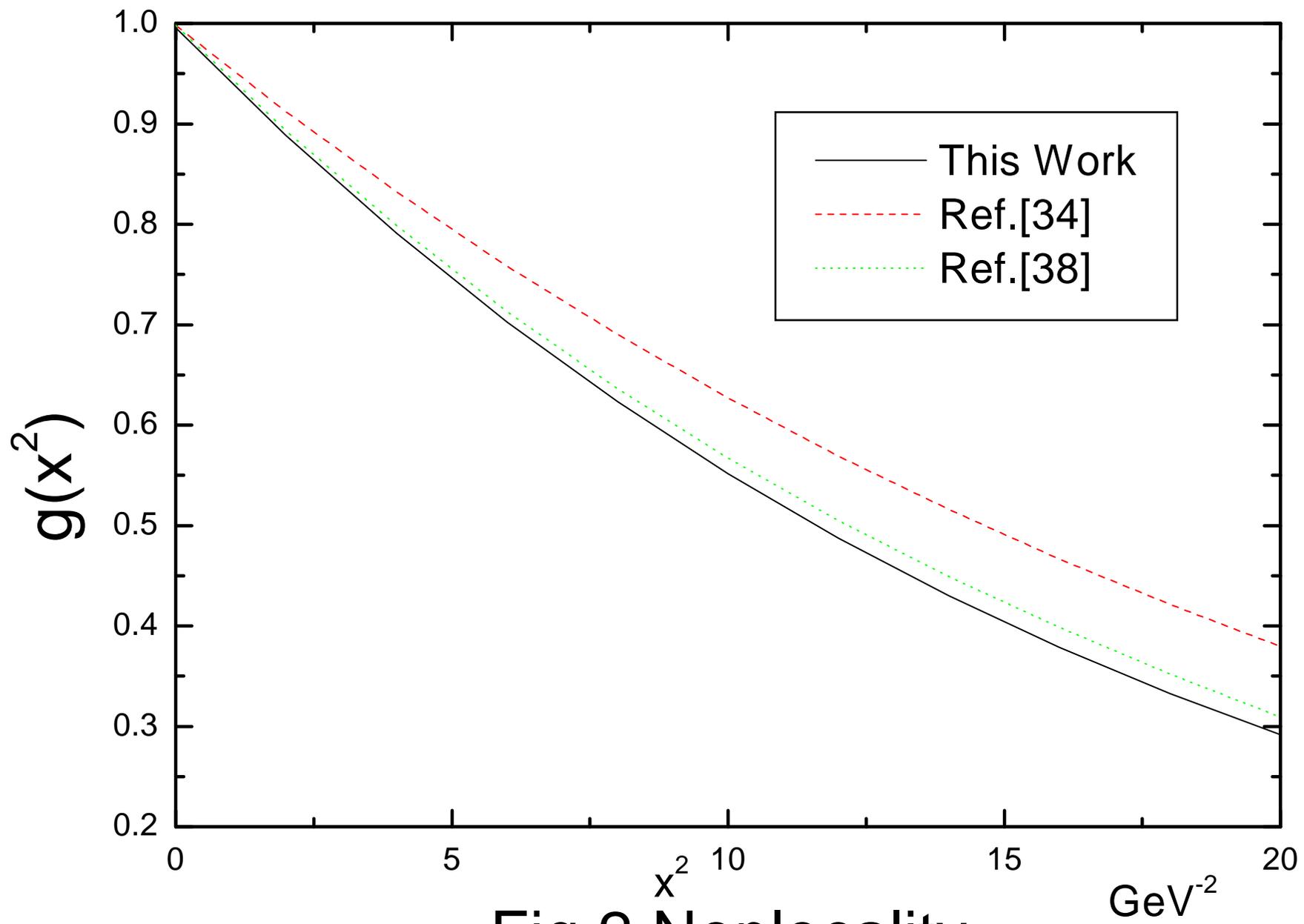
\end{document}